\documentclass[12pt]{article}
\usepackage{amssymb}

\newcommand{\hdt}{\overline{\overline{h}}}
\newcommand{\p}{\partial}

\usepackage[utf8]{inputenc}
\usepackage{hyperref}
\usepackage{amsmath,,calrsfs}
\usepackage{amsmath,amssymb}
\usepackage{graphicx}
\usepackage{calrsfs}
\usepackage{breqn}
\DeclareMathAlphabet{\pazocal}{OMS}{zplm}{m}{n}

\usepackage{tikz}
\usepackage{ragged2e}
\usepackage{indentfirst}
\usepackage[nottoc,numbib]{tocbibind}
\usepackage{booktabs, multirow} 
\usepackage{soul}
\usepackage[brazil,english]{babel}

\newcommand{\be}{\begin{equation}}

\def \del{\partial}
\newcommand{\ee}{\end{equation}}

\newcommand{\lbdd}{\overline{\Lambda}^{\ddagger}}
\newcommand{\lb}{\overline{\Lambda}}

\numberwithin{equation}{section}

\newcommand{\nn}{\nonumber}

\newcommand{\no}{\noindent}

\def\bel{\begin{align}}
\def\eel{\end{align}}
\def\bea{\begin{eqnarray}}
\def\eea{\end{eqnarray}}
\def\beq{\begin{eqnarray}}
\def\eeq{\end{eqnarray}}

\let\mathcal\pazocal
\begin{document}

\title{\textbf{The dimensional reduction of the spin-3 and spin-4 partially broken Fronsdal models}}
\author{D. Dalmazi\footnote{denis.dalmazi@unesp.br}  and A. L. R. dos Santos\footnote{alessandroluiz1905@gmail.com} \\
\textit{{ UNESP - Campus de Guaratinguet\'a - Departamento de F\'isica}}\\
\textit{{CEP 12516-410 - Guaratinguet\'a - SP - Brazil.} }\\}
\date{\today}
\maketitle

\begin{abstract}

Here we  perform a Kaluza-Klein dimensional reduction from $D+1$ to $D$ dimensions of a partially broken Fronsdal  model for spin-3 and spin-4. The reduced models describe massive spin-3 and spin-4 particles without extra propagating modes. They differ from Singh-Hagen models by quadratic terms which vanish on shell, similarly to their higher dimensional massless parents.  A covariant description of the reduced models in terms of gauge invariant field combinations naturally leads to  higher derivative models, contrary to lower spin cases ($s\le 2$).

 For both $s=3$ and $s=4$ we have a rather peculiar massless limit. Although their particle content corresponds to the $\pm j$ helicities with $j=0,1, \cdots s$, as expected, they appear intertwined in a non trivial way preventing diagonalization through any local field redefinition.

\end{abstract}
\newpage

\section{Introduction}

From the fundamental standpoint of the Poincaré symmetry, elementary massive and massless particles of arbitrary integer and half-integer spins are allowed to exist. However, so far
we have only observed spins $s=0,1/2,1$.
It is tempting to associate that experimental fact with the theoretical difficulties that one finds in trying to define self-consistent interacting Higher Spin (HS) theories, specially for $s>2$. Higher spins require higher rank tensors containing many non physical components which must be eliminated via appropriate symmetries and constraints already at free level. Those symmetries and constraints can hardly stand the introduction of interactions. Several types of problems as triviality of interactions, non closure of the gauge algebra, lack of causality and/or unitarity arise, see the review works \cite{snow,rakibur_r,sagnoti_r1,vasiliev_r,sagnoti_r}, the book \cite{book}
and the so called ``No-Go'' theorems reviewed in \cite{porrati,bbs}. 

Alternatively, one can avoid the ``No-Go'' theorems by adding  an infinite number of HS particles altogether as in string theory \cite{rakibur_r} (massive) or in the on shell formulation of Vasiliev \cite{vasiliev,vasiliev2} (massless) in $AdS$ spaces.  However, if one insists in describing each spin-s separately starting, for integer spin, from the paradigmatic Fronsdal massless model \cite{fronsdal}, it turns out that the nonlinear completion program faces severe problems, as we have mentioned, see also \cite{bulk}.

In principle, the less symmetry we require for building up the interacting model, the more choices we have for the vertices. Thus, it is natural to ask whether the local symmetries of the Fronsdal model constitute a  minimal set. The Fronsdal model is invariant under a sort of linearized HS diffeomorphism (diff henceforth): $\delta \, h = \p \overline{\Lambda}$, with traceless gauge parameters $\eta\cdot\overline{\Lambda}=0$. The authors of \cite{pbf} have proved, in the light cone gauge, that for arbitrary integer spin-s it is possible to partially break the traceless diff symmetry off-shell while still having an irreducible description of free massless spin-s particles with only $\pm s$ propagating helicities in $D=4$. The traceless gauge parameter  is further required to have vanishing double divergence: $\p\cdot\p\cdot \lbdd =0 $. Explicitly Lorentz covariant evidences have been added in \cite{epjc26} for the spin-3 and spin-4 cases where it is shown that the two-point amplitudes of the partially broken Fronsdal (PBF) model of \cite{pbf} differ from the Fronsdal ones by contact terms. This guarantees that the PBF and the Fronsdal model share the same spectrum and indicates the absence of a dual map between the two models which might hold at action level, including possible vertices, where we have the product of fields at the same point. 

Since massive models can be obtained from Kaluza-Klein (KK) dimensional reductions of massless models, it is natural to ask about how the differences between the PBF and Fronsdal models manifest in their massive reductions. This is the aim of the present work. In sections 2 and 3 we carry out the reductions of the spin-3 and spin-4 PBF models respectively and show that  the reduced models propagate only massive spin-3 and massive spin-4 particles without any further modes. We also show that the corresponding models differ from the usual Singh-Hagen \cite{sh} ones by perfect squares of field combinations which vanish on shell mirroring in a remarkable way their higher dimensional massless parents. In section 4 we investigate the curious massless limit of the reduced models which reproduce the expected helicities in a unusual non diagonal form.  We also show,
in the spin-3 case, how this is absolutely necessary to reproduce the contact terms obtained in \cite{epjc26}. In section 5 we draw our conclusions. The short appendix (section 7) is left for some details of the calculations.

\section{Dimensional reduction}

\subsection{Introduction}

The PBF model introduced in \cite{pbf} can be defined as a one-parameter deformation of the Fronsdal model. For arbitrary integer spin-s we have\footnote{We work in flat Minkowski spacetime with $\eta = \text{diag}(-,+,+,\dots,+)$ and $D\ge 3$ dimensions. Symmetrizations have no weights, e.g., $T_{(\mu\nu)} = T_{\mu\nu} + T_{\nu\mu}$. We use in part of the work the compact notation of \cite{fs1}, where symmetrizations are implicit: $\partial^k h = \partial_{(\mu_1}\cdots \partial_{\mu_k} h_{\mu_{k+1} \dots \mu_{k+s})}\,$. The dotted terms $\partial \cdot$ and $\eta \cdot$ represent contractions: $(\partial \cdot)^k h = \partial^{\mu_1} \cdots \partial^{\mu_k} h_{\mu_1 \dots \mu_k\mu_{k+1}\cdots\mu_s}$ and $(\eta \cdot)^k h = \eta^{\mu_1 \mu_2} \cdots \eta^{\mu_{2k-1}\mu_{2k}} h_{\mu_1\cdots\mu_{2k-1}\mu_{2k}\mu_{2k+1} \cdots \mu_s}$. Primes denote lower $k$ traces, e.g., $h^{'}=\eta\cdot h$ and  $h^{''}=\eta\cdot\eta\cdot h$. A single bar indicates a traceless field ($\eta^{\mu_1\mu_2} \bar{h}_{\mu_1\mu_2 \dots \mu_s} = 0$).}

\bea \pazocal{L}_{PBF}^{(s)}(c) &=&\!\!\!  -\frac{1}{2} (\partial h)^2 + \frac{s}{2}(\partial\! \cdot\! h)^2 + \frac{s(s-1)}{2}\ \partial\! \cdot\! \partial\! \cdot\! h\ h' + \frac{s(s-1)}{4}\ (\partial h')^2  \nn\\ &+& \frac{c\, s(s-1)(s-2)}{8} \, (\partial \cdot h')^2 \label{lc} \\ &=&
\pazocal{L}_{F}^{(s)} - f\, (\partial \cdot h')^2  \nn \eea

\no where $ \pazocal{L}_{F}^{(s)}= \pazocal{L}_{PBF}^{(s)}(c=1)$ is the usual Fronsdal model \cite{fronsdal} and we define the deformation parameter

\be f \equiv \frac{s(s-1)(s-2)(1-c)}8 \quad ,  \label{f} \ee

 The rank-s symmetric tensor is double traceless: $\eta\cdot\eta\cdot h=h^{\prime\prime}=0$. The action corresponding to (\ref{lc}) is invariant under traceless HS diffs with vanishing double divergence,

\be \delta h = \p \lbdd \qquad ; \qquad \p\cdot\p\cdot \lbdd =0=\eta\cdot \lbdd \label{dd} \ee

\no At $c=1$ the symmetry is enhanced to traceless HS diffs $\delta h = \p \lb $. The less symmetry we have, the more freedom is left for the source coupling. Invariance of the integral of the source term $h_{\mu_1\cdots \mu_s}T^{\mu_1\cdots \mu_s}$ under (\ref{dd})  allows the non conservation law:

\be \p \cdot T = \eta\, \Omega + \p\, \p \, J , \label{source} \ee

\no where $\Omega$ and $J$ are arbitrary rank-($s-3$) symmetric tensors constrained only by a vanishing double trace of the right hand side  of (\ref{source}) for $s \ge 5$. In the Fronsdal theory ($f=0$) we have less freedom for the sources and we need to set $J=0$. 

The fact that the PBF model describes only spin-s helicities has been demonstrated in \cite{pbf} in the light cone gauge for arbitrary integer spins. For the spin-3 and spin-4 cases one has calculated the exchange amplitude in a explicitly Lorentz covariant way in \cite{epjc26}. It is shown that they differ from the usual Fronsdal case only by analytic terms (contact terms). So it is expected that the dimensional reduction of (\ref{lc}) describes massive spin-s particles like the Singh-Hagen model \cite{sh}. In the next subsections we perform the dimensional reduction of the spin-3 case. 

\subsection{The spin-3 case}

We use capital Latin letters to denote the indices of the $(D+1)$-dimensional Minkowski space, ($M,N,\ldots=0,1,2,\ldots,D$). In  $D$-dimensions we use Greek letters ($\mu,\nu,\ldots=0,1,2,\ldots,D-1$). We compactfy the last spatial dimension $x^{D}=y$ on a circle of radius $R=1/m$ and keep only one massive mode. Our starting point is the  spin-3 PBF model  in $D+1$ dimensions,

\be S_{PBF}^{(3)}= \int{d^{D+1}x} \, {\cal L}_{PBF}^{(3)} \quad . \label{s3d+1} \ee

\no where

\begin{align}
    \mathcal{L}_{PBF}^{(3)} = & -\frac{1}{2} (\partial_M H_{NAB})^2 + \frac{3}{2} (\partial^M H_{MNA})^2 + 3 (\partial^M \partial^N H_{MNA}) H^A \nn\\
    & + \frac{3}{2} (\partial_M H_N)^2 + \left( \frac{3}{4} - f \right) (\partial^M H_M)^2 \quad , \label{lf3}
\end{align}

\no with $f=(3/4)(1-c)$. The vector field stands for the trace: $H_M=\eta^{AB}H_{ABM}$. The action is invariant under $\delta H_{MNA}=\p_{(M}\lbdd_{NA)}$. The tensor field and the gauge parameter can be decomposed as follows 
\begin{align*}
    H_{\mu\nu\alpha} &= \sqrt{\frac{m}{\pi}} h_{\mu\nu\alpha}(x) \cos(my) \\
    H_{\mu\nu D} &= \sqrt{\frac{m}{\pi}} W_{\mu\nu}(x) \sin(my) \\
    H_{\mu D D} &= \sqrt{\frac{m}{\pi}} A_\mu(x) \cos(my) \\
    H_{D D D} &= \sqrt{\frac{m}{\pi}} z(x) \sin(my)
\end{align*}

\begin{align*}
    \Lambda_{\mu\nu} &= \sqrt{\frac{m}{\pi}} \lambda_{\mu\nu}(x) \cos(my) \\
    \Lambda_{\mu D} &= \sqrt{\frac{m}{\pi}} \lambda_\mu(x) \sin(my) \\
    \Lambda_{D D} &= \sqrt{\frac{m}{\pi}} \phi(x) \cos(my)
\end{align*}

\no The traceless condition on the gauge parameter requires 

\be \eta^{MN}\lbdd_{MN}=\phi + \lambda = 0 \quad ; \quad \lambda \equiv \eta^{\mu\nu}\lambda_{\mu\nu} \quad , \label{c3a} \ee

\no while the vanishing double divergence $\p^M\p^N\lbdd_{NM}=0$ implies

\be \p^{\mu}\p^{\nu}\lambda_{\mu\nu} + 2\, m \, \p^{\mu}\lambda_{\mu} + m^2 \lambda = 0  \quad . \label{c3b}\ee

\no After eliminating $\phi$ through (\ref{c3a}), the gauge transformations in $D$-dimensions become:

\begin{align}
\delta\, h_{\mu\nu\rho} &= \p_{(\mu}\lambda_{\nu\rho)} \label{d1}\\
\delta\, W_{\mu\nu} &= \p_{(\mu}\lambda_{\nu )} - m\, \lambda_{\mu\nu} \label{d2}\\
\delta \, A_{\mu} &= - \p_{\mu} \lambda + 2\, m\, \lambda_{\mu} \label{d3}\\
\delta \, z &= 3\, m \, \lambda \label{d4}
\end{align}

\no Integrating over the extra dimension $y$ in (\ref{s3d+1}) we obtain a spin-3 massive model in $D$ dimensions whose $f$-dependent terms add up to a perfect square, namely,

\be \mathcal{L}_{f}^{(3)}(m) = \mathcal{L}_{f=0}^{(3)}(m)  - f \left(\p \cdot A + \p \cdot h  + m\, W + m\, z \right)^2 \quad . \label{lfm3} \ee

\no where 

\begin{align}
    \mathcal{L}_{f=0}^{(3)}(m) = & -\frac{1}{2} (\partial_\mu h_{\nu\rho\sigma})^2 + \frac{3}{2} (\partial^\mu h_{\mu\nu\rho})^2 + \frac{3}{2} (\partial_\mu h_\nu)^2 - 3 \partial_\mu h_\nu \partial_\alpha h^{\mu\nu\alpha} + \frac{3}{4} (\partial_\mu h^\mu)^2 \nn\\
    & - \frac{m^2}{2} h_{\mu\nu\rho}^2 + \frac{3 m^2}{2} h_\mu^2 - 3\, m h_{\mu\nu\rho} \partial^\mu W^{\nu\rho} + 6\, m h_\mu \partial_\nu W^{\mu\nu} - \frac{3}{2} \,m h_\mu \partial^\mu W \nn\\
    & + 3 h_{\mu\nu\rho} \partial^\mu \partial^\nu A^\rho - 3 h_\mu \Box A^\mu - \frac{3}{2} h_\mu \partial^\mu \partial_\nu A^\nu \nn\\
    & - \frac{3}{2} (\partial_\mu W_{\nu\rho})^2 + 3 (\partial_\mu W^{\mu\nu})^2 + \frac{3}{2} (\partial_\mu W)^2 - 3 \partial_\mu W \partial_\nu W^{\mu\nu} + \frac{9}{4} m^2 W^2 \nn\\
    & - 3 \partial_\mu W^{\mu\nu} \partial_\nu z + 3 \partial_\mu W \partial^\mu z - \frac{9}{2} m \,W \partial_\mu A^\mu + \frac{3}{2} m^2 W z \nn\\
    & + \frac{9}{4} (\partial_\mu A^\mu)^2 + \frac{3}{2} m\, A_\mu \partial^\mu z + (\partial_\mu z)^2 + \frac{1}{4} m^2 z^2 \label{lf3m}
\end{align}

 The Lagrangian (\ref{lf3m}) is the massive model that one obtains from the dimensional reduction of the spin-3 Fronsdal model without any gauge fixing while (\ref{lfm3}) is invariant under (\ref{d1})-(\ref{d4}) constrained by (\ref{c3b}).
 
 If $f=0$ the symmetry increases, $\lbdd \to \lb$, and the constraint (\ref{c3b}) is removed. The mass dependent terms on the right hand side of (\ref{d2})-(\ref{d4}) are Stueckelberg like gauge transformations. It is natural to proceed further by  eliminating the pure gauge Stueckelberg fields at action level. However, not all fields $(W_{\mu\nu},A_{\mu},z)$ are pure gauge. The unconstrained gauge parameters $(\lambda_{\mu\nu},\lambda_{\mu})$ are not enough to remove them. There are subtleties in the gauge fixing procedure, even at $f=0$. For instance, one might think of fixing $W_{\mu\nu}=0=A_{\mu}$ but this is not allowed, since once we get rid of $W_{\mu\nu}$ using all components of $\lambda_{\mu\nu}$ we have no freedom to move the longitudinal component of the vector field because $\delta(\p\cdot A - m \, W)=(m^2-\Box)\lambda$ and the trace $\lambda$ has been already used. On the other hand, there is no contradiction in fixing the traceless part of the rank-2 tensor to zero altogether with the vector field and $z$, namely: ${\overline W}_{\mu\nu}=0=z=A_{\mu} $. This is the gauge used in \cite{ady} and \cite{raki} which leads to the spin-3 Singh-Hagen (SH) model $ \mathcal{L}_{f=0}^{(3)}=  \mathcal{L}_{f=0}^{(3)}(h_{\mu\nu\rho},W)$ whose equations of motion imply $W=0$ and the Klein-Gordon equations $(\Box - m^2)h_{\mu\nu\rho}=0$ alltogether with the Fierz-Pauli (FP) constraints $h_{\mu}=0=(\p\cdot h)_{\mu\nu}$. So in $D=4$, we have the expected $2\, s +1 = 7$ propagating massive modes.

 At this point, regarding the general case $f\ne 0$, before discussing possible gauge choices and their subtleties, we have found more enlightening to first implement an explicitly gauge invariant approach that we have put forward in \cite{da-hs,nstd,pbf} as we explain in the next subsection. The idea is to rewrite the reduced model (\ref{lfm3})  in terms of a set of independent gauge invariants that can be derived systematically from the gauge transformations. This opens the way to embrace higher derivative massive models which would otherwise stem only from the dimensional reduction of higher derivative higher spin massless models as those of \cite{fs1} and \cite{more}.
 
 \subsection{Gauge invariants from gauge transformations}

 Let us start with simpler examples. In the spin-1 case, the dimensional reduction of the Maxwell theory leads to the Proca theory with a scalar Stueckelberg field ${\cal L}_P (A_{\mu},\varphi)$. From the $D+1$ equations $(\delta\, A_{\mu},\delta\, \varphi)=(\p_{\mu}\lambda, m\, \lambda)$ we can eliminate the gauge parameter $\lambda = \delta\, 
 \varphi/m$ and plugging back in the remaining $D$ equations we have a vector invariant $  \delta \, I_{\mu}  \equiv \delta\, ( A_{\mu}- \p_{\mu}\varphi/m) = 0 $. The Proca model can be written in an explicit gauge invariant way ${\cal L}_P (A_{\mu},\varphi)={\cal L}_P (I_{\mu},0)$. Analogously, in the dimensional reduction of the spin-2 linearized Einstein-Hilbert (LEH) theory we have $D(D+1)/2 + D+1$ equations

 \be
\delta\, h_{\mu\nu} = \p_{(\mu}\, \lambda_{\nu)} \quad ; \quad 
\delta \, A_{\mu} = \p_{\mu}\, \lambda - m\, \lambda_{\mu} \quad ; \quad
\delta\, \varphi = 2\, m\, \lambda \label{dphi} 
 \ee

 \no If we take $D+1$ of the previous equations  to eliminate  $(\lambda_{\mu},\lambda)$ in terms of field variations and plug back in the remaining gauge transformations we end up with $D(D+1)/2$ invariants,

\be \delta \, I_{\mu\nu} \equiv \delta \left\lbrack h_{\mu\nu} + \frac{\p_{(\mu}\, A_{\nu)}}m - \frac{\p_{\mu}\p_{\nu}\varphi}{m^2}\right\rbrack = 0 \quad . \label{imn}\ee

\no The resulting massive spin-2 FP model can be   entirely written in terms of those invariants:  ${\cal L}_{FP}(h_{\mu\nu},A_{\mu},\varphi)= {\cal L}_{FP}(I_{\mu\nu},0,0)$. The would be higher derivatives cancel out.

From the previous examples we learn that the number of gauge invariants is the difference between the total number of fields and the number of independent gauge parameters, see also \cite{nstd,pbf}, 

\be N_I = N_{\Phi} - N_{\lambda} \quad . \label{ni} \ee

If we apply (\ref{ni}) to the reduced massive spin-3 model (\ref{lf3m}) with $f=0$, given the fields $\Phi = (h_{\mu\nu\rho},W_{\mu\nu},A_{\mu},z)$ and the independent gauge parameters are $(\lambda_{\mu\nu},\lambda_{\mu})$, we expect a totally symmetric rank-3 tensor and a scalar as gauge invariants: $(I_{\mu\nu\rho},\mathbb{I})$, see also  \cite{raki}. As we move to the PBF model $f\ne 0$, since we have one less independent gauge parameter due to the double divergence constraint (\ref{c3b}), we expect one new scalar invariant: $\mathbb{I}_N$. Indeed, repeating the procedure of the spin-1 and spin-2 cases, if we use (\ref{d1})-(\ref{d4}) to eliminate the gauge parameters in terms of the field transformations and plug them back in the remaining equations and in the constraint (\ref{c3b}) we obtain the independent gauge invariants:

\begin{equation}
    I_{\mu\nu\alpha} = h_{\mu\nu\alpha} + \frac{\partial_{(\mu} W_{\nu\alpha)}}{m} - \frac{\partial_{(\mu} \partial_\nu A_{\alpha)}}{m^2} - \frac{\partial_\mu \partial_\nu \partial_\alpha z}{m^3}  \quad , \label{imna}
\end{equation}

\begin{equation}
    \mathbb{I} = 3\,W - \frac{3\,\partial_\mu A^\mu}{m} - \frac{(\Box - m^2)z}{m^2} \quad , \label{i}
\end{equation}

\begin{equation}
    \mathbb{I}_N = - \frac{3\, \p^{\mu}\p^{\nu}W_{\mu\nu}}{m^2} + \frac{3(\Box + m^2)\, \p^{\mu}A_{\mu}}{m^3}+\frac{\left\lbrack \Box(\Box + m^2)+m^4\right\rbrack z}{m^4} \quad , \label{in}
\end{equation}

\no In terms of which, the reduced PBF model (\ref{lfm3}) becomes

\be \mathcal{L}_{f}(I_{\mu\nu\alpha},\mathbb{I},\mathbb{I}_N) = \mathcal{L}_{f=0}(I_{\mu\nu\alpha},\mathbb{I})  - \frac 49\, m^2 f \left\lbrack  \mathbb{I}_N - 
\frac{(\Box - m^2)\mathbb{I}}{2\,m^2} + \frac{3\,\p^{\mu}I_{\mu}}{2\,m}\right\rbrack^2 \quad . \label{lfm3i} \ee

\no where $ I_{\mu} = \eta^{\nu\alpha}I_{\mu\nu\alpha}$ and the reduced Fronsdal model (\ref{lf3m}) is written as a fourth order model,

\begin{align}
    \mathcal{L}_{f=0}(I_{\mu\nu\alpha},\mathbb{I}) = & -\frac{1}{2} (\partial_\mu I_{\nu\alpha\beta})^2 + \frac{3}{2} (\partial_\mu I^{\mu\nu\alpha})^2 + \frac{3}{2} (\partial_\mu I_\nu)^2 - 3 \partial_\mu I_{\nu} \partial_\alpha I^{\mu\nu\alpha} + \frac{3}{4} (\partial_\mu I^\mu)^2 \nn\\
    & -\frac{m^2}{2} I_{\mu\nu\alpha}^2 + \frac{3}{2} m^2 I_\mu^2 + \frac{\partial_\mu I^{\mu\nu\alpha} \partial_\nu \partial_\alpha \mathbb{I}}{m} - \frac{3}{2} \frac{\partial_\mu I^\mu \Box \,\mathbb{I}}{m} \nn\\
    & - \frac 32\,m \, I_\mu \,\partial^\mu \mathbb{I} +  \frac{(\Box\, \mathbb{I})^2}{4\,m^2} + \frac{1}{2} (\partial_\mu \mathbb{I})^2 + \frac{m^2}{4}  \mathbb{I}^2  \quad , \label{l34th}
\end{align}

\no Since linear combinations of gauge invariants are also gauge invariants, we can always redefine them. In particular, after the following redefinitions

\be I_{\mu\nu\alpha} = \tilde{I}_{\mu\nu\alpha} + \frac1{3\, m\, D}\eta_{(\mu\nu}\p_{\alpha)} \mathbb{I} \quad , \label{i3til} \ee

\be  \mathbb{I}_N = \tilde{\mathbb{I}}_N - \frac{(2\Box + m^2 D)\mathbb{I}}{2\, m^2\, D}   - \frac{3\,\p^{\mu}\tilde{I}_{\mu}}{2\, m} \quad , \label{intil} \ee 

\no we get rid of fourth order terms and  obtain the simpler second order model

\be \mathcal{L}_{f}(\tilde{I}_{\mu\nu\alpha},\mathbb{I},\tilde{\mathbb{I}}_N) = \mathcal{L}_{SH}(\tilde{I}_{\mu\nu\alpha},\mathbb{I})  - 4\, m^2 f \,\tilde{\mathbb{I}}_N^2. \label{L2} \ee

\no where $ \mathcal{L}_{SH}(\tilde{I}_{\mu\nu\alpha},\mathbb{I})$ is  the usual spin-3 Singh-Hagen  model \cite{sh} in terms of gauge invariants:

\begin{align}
    \mathcal{L}_{SH}(\tilde{I}_{\mu\nu\alpha},\mathbb{I}) = & -\frac{1}{2} (\partial_\mu \tilde{I}_{\nu\alpha\beta})^2 + \frac{3}{2} (\partial_\mu \tilde{I}^{\mu\nu\alpha})^2 + \frac{3}{2} (\partial_\mu \tilde{I}_\nu)^2 - 3 \partial_\mu \tilde{I}_{\nu}\partial_ \alpha  \tilde{I}^{\mu\nu\alpha} + \frac{3}{4} (\partial_\mu \tilde{I}^\mu)^2 \nn\\    & -\frac{m^2}{2} \tilde{I}_{\mu\nu\alpha}^2 + \frac{3\,m^2}{2} m^2 \tilde{I}_\mu^2 - \frac{m\,(D-2)}{2D} \tilde{I}_\mu \partial^\mu \mathbb{I} + \frac{(D-1)(D-2)}{6D^2} (\partial_\mu \mathbb{I})^2 + \frac{m^2}{4} \mathbb{I}^2  \label{lsh}
\end{align}

  If we could consider the independent gauge invariants $(\tilde{I}_{\mu\nu\alpha},\mathbb{I},\tilde{\mathbb{I}}_N)$ as fundamental fields,  we could say that  (\ref{L2}) and $\mathcal{L}_{SH}$  differ by a trivial (non dynamic) square term which assures that both models describe physical massive spin-3 particles without extra modes. In the next subsection we fix convenient gauges at action level and  prove, in terms of fundamental fields, that the previous interpretation is indeed correct. 

\subsection{Unitary gauges and particle content}
 
Now in order to check the content of (\ref{lfm3}) we return to the representation in terms of the original fields by fixing unitary gauges. According to \cite{moto}, all gauges that can be achieved by a unique choice of the gauge parameters can be fixed at action level without altering the particle content of the theory. In practice this amounts to gauge away field combinations whose gauge transformations do not depend on derivatives. The best we can do is to get rid of the rank-($s-1$) Stueckelberg field. In the spin-1 proca theory we can safely set $\varphi =0$. In the spin-2 case it corresponds to fix $\varphi_{\mu}\equiv A_{\mu} - \p_{\mu}\phi/(2m)=0$ since $\delta\, \varphi_{\mu}=-m\lambda_{\mu}$. Such gauge eliminates both Stueckelberg fields $(A_\mu,\phi)$ at once and $I_{\mu\nu}\to h_{\mu\nu}$. We end up with the usual FP massive spin-2 theory solely in terms of $h_{\mu\nu}$. The spin-3 analogue of those unitary gauges is given by

\be \varphi_{\mu\nu} \equiv W_{\mu\nu} - \frac{\p_{(\mu}A_{\nu)}}{2\, m} - \frac{\p_{\mu}\p_{\nu}z}{3\, m^2} = 0 \quad , \label{fmn} \ee 

\no since from (\ref{d2})-(\ref{d4}) we have the pure gauge transformation: $\delta \varphi_{\mu\nu} = -m\, \lambda_{\mu\nu}$. . In the spin-3 case however, it is known that we can not eliminate all auxiliary fields.  At the gauge (\ref{fmn}) we have

\be I_{\mu\nu\alpha} = h_{\mu\nu\alpha} \quad ; \quad  \mathbb{I}=z \quad ; \quad \mathbb{I}_N =\frac 3m \p\cdot A + \frac{(\Box + m^2)z}{m^2}\label{gfinv3} \ee

\no Now the reduced PBF model (\ref{lfm3i}) can be written as 

\be
{\cal L}_{f}(h,A,z) = {\cal L}_{f=0}(h,z) - 4\, f \left\lbrack \p \cdot A + \frac{(\Box + 3\,m^2)z}{6\, m} + \frac{\p^{\mu}h_{\mu}}{2} \right\rbrack^2  \quad . \label{lint3} \ee

\no where

\bea  
{\cal L}_{f=0}(h,z) &=& {\cal L}_F^{(3)}(h_{\mu\nu\alpha}) + \frac{m^2}2\left(3\, h_\mu^2- h_{\mu\nu\alpha}^2\right) + \frac{h^{\mu\nu\alpha}\p_{\mu}\p_{\nu}\p_{\alpha}\, z}m + \frac 3{2\, m}(\p^{\mu}h_{\mu}) \, \Box \, z \nn \\
&-& \frac {3\, m}2\, z \,\p^{\mu}h_{\mu}+ \frac{(\Box\, z)^2}{4\, m^2} - \frac 12 z\, \Box \, z + \frac{m^2\, z^2}4 \label{lhz4} 
\eea

 In order to figure out the particle content of the reduced PBF model (\ref{lint3}), the key point is to notice that the vector field $A_{\mu}$ only appears in the second term (perfect square). Its equations of motion $\p_{\mu} \left\lbrack \p \cdot A + g(z,h) \right\rbrack =0 $, assuming vanishing fields at infinity, amount to $\left\lbrack \p \cdot A + g(z,h) \right\rbrack =0 $,  independently of  whatever $g(z,h)$ is. The contributions of $g(z,h)$ to $z$ and $h_{\mu\nu\alpha}$ equations of motion will be proportional to $\left\lbrack \p \cdot A + g(z,h) \right\rbrack  $ which vanishes on shell.   Thus, the $f$ dependent perfect square in (\ref{lint3}) gives no on shell contribution whatsoever and the spin-3 reduced PBF model has the same content of (\ref{lhz4}), i.e., the reduced Fronsdal ($f=0$) model in the gauge\footnote{Notice that the higher derivatives, not present in (\ref{lf3m}), have been introduced by the gauge (\ref{fmn}).}  (\ref{fmn}). After an invertible field redefinition,

\be h_{\mu\nu\alpha} \to \tilde{h}_{\mu\nu\alpha} + \frac 1{3\, D\,m}\eta_{(\mu\nu}\p_{\alpha)} \, z \quad , \label{hredef} \ee 

\no we get rid of all higher derivative terms in (\ref{lhz4}) and recover the second order spin-3 Singh-Hagen model (\ref{lsh}) with the replacements $(\tilde{I}_{\mu\nu\alpha}, \mathbb{I}) \to (\tilde{h}_{\mu\nu\alpha}, z)$. So the content of (\ref{lint3}) or (\ref{lhz4}) corresponds to only  massive spin-3 particles.

 It is worth mentioning that (\ref{lhz4}), after the redefinition  $z \to m\, z$, is the dimensional reduction of the massless spin-3 fourth order model of Francia and Sagnotti, see \cite{fs1}, after a gauge fixing. The equations of motion of the model of \cite{fs1} had been previously obtained by J. Schwinger \cite{schw}. The massless limit of (\ref{lhz4}), after  $z \to m\, z$, is invariant under unconstrained diffs where $z$ works like a Stueckelberg field, see \cite{fs1}.
 
Next we turn to the dimensional reduction of the spin-4 PBF model.

\section{Dimensional reduction of the spin-4 PBF model}

Now we start with

\be S_{PBF}^{(4)}= \int{d^{D+1}x} \, {\cal L}_{PBF}^{(4)} \quad , \label{s4d+1} \ee

\no where

\begin{align}
    \mathcal{L}_{PBF}^{(4)} = & -\frac{1}{2} (\partial_M H_{NABC})^2 + 2\, (\partial^M H_{MNAB})^2 + 6\, H_{MNAB}\partial^M \partial^N H^{AB} \nn\\
    & + 3\, (\partial_M H_{AB})^2 + \left(3 - f \right) (\partial^M H_{MN})^2 \quad . \label{lf4}
\end{align}

\no At $s=4$ we have $f=3\, (1-c)$ and $H_{AB}= \eta^{MN}H_{MNAB}$. The $D+1$ decomposition of the double traceless field $H_{MNAB}$ is given by

\begin{align*}
    H_{\mu\nu\alpha\beta} &= \sqrt{\frac{m}{\pi}} h_{\mu\nu\alpha\beta}(x) \cos(my) \quad,\\
    H_{D\mu\nu\alpha} &= \sqrt{\frac{m}{\pi}} \chi_{\mu\nu\alpha}(x) \sin(my) \quad,\\
    H_{DD\mu\nu} &= \sqrt{\frac{m}{\pi}} W_{\mu\nu}(x) \cos(my) \quad,\\
    H_{DDD\mu} &= \sqrt{\frac{m}{\pi}} B_{\mu}(x) \sin(my) \quad,\\
    H_{DDDD} &= \sqrt{\frac{m}{\pi}} z (x)\cos(my) \quad.
\end{align*}

The double traceless constraint allows us to eliminate the field $z$,

\be H^{\prime\prime} = \eta^{AB}\eta^{CD}H_{ABCD} = z + 2\, W +h =0 \label{zfield} \quad.  \ee

The rank-3  gauge parameter can be decomposed as 

\begin{align*}
\Lambda_{\mu\nu\alpha} &= \sqrt{\frac{m}{\pi}} \Lambda_{\mu\nu\alpha}(x) \cos(my) \quad,\\
\Lambda_{D\mu\nu} &= \sqrt{\frac{m}{\pi}} \Sigma_{\mu\nu}(x) \sin(my) \quad,\\
\Lambda_{DD\mu} &= \sqrt{\frac{m}{\pi}} \xi_{\mu}(x) \cos(my) \quad,\\
\Lambda_{DDD} &= \sqrt{\frac{m}{\pi}} \lambda(x) \sin(my) \quad.
\end{align*}

\noindent The parameters $\xi_{\mu}$ and $\lambda$ can be eliminated by the   traceless conditions $\Lambda_{M}=0$ :
\begin{equation}
\Lambda_{\mu} + \xi_{\mu} = 0, \quad \Sigma + \lambda = 0 \quad , \label{trss4}
\end{equation}

\no while the restrictions $\p^{M} \p^{N} \Lambda_{MNA} =0$ lead to the constraints

\begin{equation}
\partial^{\alpha}\partial^{\beta}\Lambda_{\mu\alpha\beta} + 2m\partial^{\nu}\Sigma_{\mu\nu} + m^{2}\Lambda_{\mu} = 0, \quad \partial^{\mu}\partial^{\nu}\Sigma_{\mu\nu} + 2m\partial^{\mu}\Lambda_{\mu} + m^{2}\Sigma = 0 \quad . \label{c3}
\end{equation}


After eliminating $z$, the reduced massive spin-4 model in $D$ dimensions is given by

\begin{equation}
\begin{aligned}
\mathcal{L}^{(4)} = & -\frac{1}{2}(\partial_{\mu}h_{\nu\alpha\beta\rho})^{2} + 2(\partial_{\mu}h^{\mu\nu\alpha\beta})^{2}
+ 6h_{\mu\nu\alpha\beta}\partial^{\mu}\partial^{\nu}h^{\alpha\beta} +3(\partial_{\mu}h_{\nu\alpha})^{2} + 3(\partial_{\mu}h^{\mu\nu})^{2} + \frac{5}{2}(\partial_{\mu}h)^{2} \\
& -\frac{m^{2}}{2}h_{\mu\nu\alpha\beta}^{2} + 3m^{2}h_{\mu\nu}^{2} + \frac{3}{2}m^{2}h^{2} - 4m\,h_{\mu\nu\alpha\beta}\partial^{\mu}\chi^{\nu\alpha\beta} + 12m\,h_{\mu\nu}\partial_{\alpha}\chi^{\mu\nu\alpha} \\
& - 6m\,h_{\mu\nu}\partial^{\mu}\chi^{\nu} + 6m\,h\partial_{\mu}\chi^{\mu} + 6h_{\mu\nu\alpha\beta}\partial^{\mu}\partial^{\nu}W^{\alpha\beta} - 6h_{\mu\nu}\Box W^{\mu\nu} - 6h_{\mu\nu}\partial^{\mu}\partial_{\alpha}W^{\alpha\nu} \\
& - 6h\partial_{\mu}\partial_{\nu}W^{\mu\nu} - 4h\Box W - 6m\,h_{\mu\nu}\partial^{\mu}B^{\nu} - 2m\,h\partial_{\mu}B^{\mu} \\
& - 2(\partial_{\mu}\chi_{\nu\alpha\beta})^{2} + 6(\partial_{\mu}\chi^{\mu\nu\alpha})^{2} + 12\chi_{\mu\nu\alpha}\partial^{\mu}\partial^{\nu}\chi^{\alpha}
 + 6(\partial_{\mu}\chi_{\nu})^{2} + 3(\partial_{\mu}\chi^{\mu})^{2} + 9m^{2}\chi_{\mu}^{2} \\
& - 18m\,\chi_{\mu}\partial_{\nu}W^{\mu\nu} - 6m\,\chi_{\mu}\partial^{\mu}W + 12\chi_{\mu\nu\alpha}\partial^{\mu}\partial^{\nu}B^{\alpha} - 12\chi_{\mu}\Box B^{\mu} - 6\chi_{\mu}\partial^{\mu}\partial_{\nu}B^{\nu} + 6m^{2}\chi_{\mu}B^{\mu} \\
& + 9(\partial_{\mu}W^{\mu\nu})^{2} + 6\partial_{\mu}W\partial_{\nu}W^{\mu\nu} + (\partial_{\mu}W)^{2} + 6m\,W_{\mu\nu}\partial^{\mu}B^{\nu} + 2m\,W\partial_{\mu}B^{\mu} \\
& + 4(\partial_{\mu}B_{\nu})^{2} + 5(\partial_{\mu}B^{\mu})^{2} + m^{2}B_{\mu}^{2} \\
& - fm^{2}(V_{\mu}^{2} + S^{2}) \quad, \label{ms4}
\end{aligned}
\end{equation}

\noindent where $(h_{\mu\nu},h)=(\eta^{\alpha\beta}h_{\alpha\beta\mu\nu},\eta^{\mu\nu}h_{\mu\nu})$ and

\begin{equation}
V_{\mu} = \frac{\partial^{\nu}W_{\mu\nu}}{m} + \frac{\partial^{\nu}h_{\mu\nu}}{m} + \chi_{\mu} + B_{\mu} \quad,
\end{equation}

\begin{equation}
S = \frac{\partial_{\mu}\chi^{\mu}}{m} + \frac{\partial_{\mu}B^{\mu}}{m} + h + W \quad.
\end{equation}

\no The model (\ref{ms4}) is invariant under

\begin{align}
    \delta h_{\mu\nu\alpha\beta} &= \partial_{(\mu} \Lambda_{\nu\alpha\beta)}  \quad ; \quad 
    \delta \chi_{\mu\nu\alpha} = \partial_{(\mu} \Sigma_{\nu\alpha)}  - m \Lambda_{\mu\nu\alpha} \quad ,  \label{gtxs4} \\
    \delta W_{\mu\nu} &= -\partial_{(\mu} \Lambda_{\nu)}  + 2\,m \,\Sigma_{\mu\nu} \quad ; \quad 
    \delta B_{\mu} = -\partial_{\mu} \Sigma + 3m \,\Lambda_{\mu} \quad , \label{gtbs4}
\end{align}

\no once the constraints (\ref{c3}) are considered. From the previous gauge transformations and constraints we constructively build up the gauge invariants:

\begin{equation}
I_{\mu\nu\alpha\beta} = h_{\mu\nu\alpha\beta} + \frac{\partial_{(\mu}\chi_{\nu\alpha\beta)}}{m} - \frac{\partial_{(\mu}\partial_{\nu}W_{\alpha\beta)}}{m^{2}} - \frac{\partial_{(\mu}\partial_{\nu}\partial_{\alpha}B_{\beta)}}{m^{3}} - \frac{\partial_{\mu}\partial_{\nu}\partial_{\alpha}\partial_{\beta}(h+2W)}{m^{4}} \;, \, \label{I4}
\end{equation}

\begin{equation}
M_{\mu} = \chi_{\mu} + \frac{B_{\mu}}{3} + \frac{\partial_{\mu}h}{12m} - \frac{\partial_{\mu}W}{3m} - \frac{\partial^{\nu}W_{\mu\nu}}{m} - \frac{\Box B_{\mu}}{3m^{2}} - \frac{2\partial_{\mu}\partial_{\nu}B^{\nu}}{3m^{2}} - \frac{\Box\partial_{\mu}(h+2W)}{4m^{3}} \;, \, \label{M4}
\end{equation}

\begin{equation}
\begin{aligned}
N_{\mu} = {}& \frac{\partial^{\alpha}\partial^{\beta}\chi_{\mu\alpha\beta}}{m^{2}} - \frac{(\Box+m^{2})}{m^{3}}\partial^{\nu}W_{\mu\nu} - \frac{\partial_{\mu}\partial_{\alpha}\partial_{\beta}W^{\alpha\beta}}{2m^{3}} - \frac{(\Box^{2}+m^{2}\Box+m^{4})}{3m^{4}}B_{\mu} \\
& - \frac{(2\Box+m^{2})}{3m^{4}}\partial_{\mu}\partial_{\nu}B^{\nu} - \frac{(3\Box^{2}+2m^{2}\Box+m^{4})}{12m^{5}}\partial_{\mu}(h+2W) \;, \, \label{N4}
\end{aligned}
\end{equation}

\begin{equation}
L = \frac{\partial^{\mu}\partial^{\nu}W_{\mu\nu}}{m^{2}} + \frac{2(\Box+2m^{2})}{3m^{3}}\partial_{\mu}B^{\mu} + \frac{(\Box^{2}+2m^{2}\Box+3m^{4})}{6m^{4}}(h+2W) \;. \, \label{L4}
\end{equation}

Notice that the constraints (\ref{c3}) are responsible for the  invariants $(L,N_{\mu})$ which only appear in the PBF model ($f\ne 0$). In terms of the gauge invariants, the reduced model can be written as 

\begin{equation}
\begin{aligned}
\mathcal{L}^{(4)}_f = {}& -\frac{1}{2}(\partial_{\mu}I_{\nu\alpha\beta\rho})^{2} + 2(\partial_{\mu}I^{\mu\nu\alpha\beta})^{2} + 6 I_{\mu\nu\alpha\beta}\partial^{\mu}\partial^{\nu}I_{\alpha\beta}
 + 3(\partial_{\mu}I_{\nu\alpha})^{2}  + 3(\partial_{\mu}I^{\mu\nu})^{2} \\
& +\frac{3}{2}I_{\mu\nu}\partial^{\mu}\partial^{\nu}I + \frac{33}{16}(\partial_{\mu}I)^{2} - \frac{m^{2}}{2}I_{\mu\nu\alpha\beta}^{2} + 3m^{2}I_{\mu\nu}^{2} + \frac{3}{2}m^{2}I^{2} + \frac{9}{16}\frac{(\Box I)^{2}}{m^{2}} \\
& - 12\frac{I_{\mu\nu\alpha\beta}\partial^{\mu}\partial^{\nu}\partial^{\alpha}M^{\beta}}{m} + 18\frac{I_{\mu\nu}\Box\partial^{\mu}M^{\nu}}{m} + 9\frac{I\Box\partial_{\mu}M^{\mu}}{m} \\
& - 18m\,I_{\mu\nu}\partial^{\mu}M^{\nu} - \frac{9}{2}m\,I\partial_{\mu}M^{\mu} - \frac{9}{2}\frac{I\Box^{2}\partial_{\mu}M^{\mu}}{m^{3}} \\
& + 9\frac{(\Box M_{\mu})^{2}}{m^{2}} + 18\frac{(\partial_{\mu}\partial_{\nu}M^{\nu})^{2}}{m^{2}} + 18(\partial_{\mu}M_{\nu})^{2} + 9(\partial_{\mu}M^{\mu})^{2} + 9m^{2}M_{\mu}^{2} + 9\frac{(\Box\partial_{\mu}M^{\mu})^{2}}{m^{4}} \\
& - fm^{2}(V_{\mu}^{2} + S^{2}) \quad, \label{6th}
\end{aligned}
\end{equation}


\no where

\begin{equation}
V_{\mu} = \frac{\partial^{\nu}I_{\mu\nu}}{m} - \frac{\partial_{\mu}I}{4m} - \frac{(\Box-m^{2})}{m^{2}}M_{\mu} - 2N_{\mu} \quad , \label{V}
\end{equation}

\begin{equation}
S = -\frac{(\Box-2m^{2})}{4m^{2}}I + \frac{(\Box-m^{2})}{m^{3}}\partial_{\mu}M^{\mu} + L \quad . \label{S}
\end{equation}


As in the spin-3 case we remark that the apparent higher derivative terms in (\ref{6th}) cancel out as we go back to the original fields, see (\ref{ms4}). Moreover, since we have the combination $\varphi_{\mu\nu\alpha}$ below transforming as pure gauge, 

\begin{equation}
\delta \varphi_{\mu\nu\alpha} \equiv \delta \left[ \chi_{\mu\nu\alpha} - \frac{\partial_{(\mu}W_{\nu\alpha)}}{2m} - \frac{\partial_{(\mu}\partial_{\nu}B_{\alpha)}}{3m^{2}} - \frac{\partial_{\mu}\partial_{\nu}\partial_{\alpha}(h+2W)}{4m^{3}} \right] = -m\, \Lambda_{\mu\nu\alpha} \, , \label{fmnr}
\end{equation}

\no if we consider (\ref{c3}) as constraints on $\Sigma_{\mu\nu}$, we can use all components of $\Lambda_{\mu\nu\rho} $ in order to fix the following gauge at action level in (\ref{6th}),

\be \varphi_{\mu\nu\rho} = 0 \quad . \label{fzero} \ee 

\no The gauge invariants become:

\begin{equation}
    I_{\mu\nu\rho} = h_{\mu\nu\rho} \quad ; \quad M_\mu = \frac{A_\mu}{3} \, , \label{im}
\end{equation}

\begin{equation}
    N_\mu = -\frac{(\partial \cdot W)_\mu}{m} - \frac{[\partial_\mu (\partial \cdot A) + (\Box + m^2) A_\mu]}{3m^2} \, , \label{nmu}
\end{equation}

\begin{equation}
    L = \frac{\partial \cdot \partial \cdot W}{m^2} + W + \frac{2(\Box + 2m^2)(\partial \cdot A)}{3m^3} + \frac{h}{2}  \, , \label{L}
\end{equation}

\no where we have redefined the vector field:
\begin{equation}
    A_\mu \equiv B_\mu + \frac{1}{4m} \partial_\mu (h + 2W) \quad . \label{30}
\end{equation}

\no In the gauge (\ref{fzero}) the Lagrangian (\ref{6th}) becomes 

\begin{equation}
\begin{aligned}
\mathcal{L}^{(4)}_f = {}& -\frac{1}{2}(\partial_{\mu}h_{\nu\alpha\beta\rho})^{2} + 2(\partial_{\mu}h^{\mu\nu\alpha\beta})^{2} + 6h_{\mu\nu\alpha\beta}\partial^{\mu}\partial^{\nu}h^{\alpha\beta}
+ 3(\partial_{\mu}h_{\nu\alpha})^{2}  + 3(\partial_{\mu}h^{\mu\nu})^{2} \\
& +\frac{3}{2}\partial^{\mu}\partial^{\nu}h + \frac{33}{16}(\partial_{\mu}h)^{2} - \frac{m^{2}}{2}h_{\mu\nu\alpha\beta}^{2} + 3m^{2}h_{\mu\nu}^{2} + \frac{3}{2}m^{2}h^{2} + \frac{9}{16}\frac{(\Box h)^{2}}{m^{2}} \\
& - 4\frac{h_{\mu\nu\alpha\beta}\partial^{\mu}\partial^{\nu}\partial^{\alpha}A^{\beta}}{m} + 6\frac{h_{\mu\nu}\Box\partial^{\mu}A^{\nu}}{m} + 3\frac{h\Box\partial_{\mu}A^{\mu}}{m} \\
& - 6m\,h_{\mu\nu}\partial^{\mu}A^{\nu} - \frac{3}{2}m\,h\partial_{\mu}A^{\mu} - \frac{3}{2}\frac{h\Box^{2}\partial_{\mu}A^{\mu}}{m^{3}} \\
& + \frac{(\Box A_{\mu})^{2}}{m^{2}} + 2\frac{(\partial_{\mu}\partial_{\nu}A^{\nu})^{2}}{m^{2}} + 2(\partial_{\mu}A_{\nu})^{2} + (\partial_{\mu}A^{\mu})^{2} + m^{2}A_{\mu}^{2} + \frac{(\Box\partial_{\mu}A^{\mu})^{2}}{m^{4}} \\
& - fm^{2}(V_{\mu}^{2} + S^{2}) \quad , \label{6thb}
\end{aligned}
\end{equation}


\noindent where $V_{\mu}$ and $S$ are given by

\begin{equation}
V_{\mu} = 2\,\frac{\partial^{\nu}W_{\mu\nu}}{m} + \frac{\partial^{\nu}h_{\mu\nu}}{m} - \frac{\partial_{\mu}h}{4m} + \frac{(\Box+3m^{2})A_{\mu}}{3m^{2}} + \frac{2}{3}\frac{\partial_{\mu}\partial_{\nu}A^{\nu}}{m^{2}} \quad,
\end{equation}

\begin{equation}
S = \frac{\partial^{\mu}\partial^{\nu}W_{\mu\nu}}{m^{2}} + W - \frac{(\Box-4m^{2})}{4m^{2}}h + \frac{(\Box+m^{2})}{m^{3}}\partial_{\mu}A^{\mu} \quad.
\end{equation}

\no Since  $W_{\mu\nu}$ only appears in $\mathcal{L}^{(6)}_f $ through $S$ and $V_{\mu}$, its equation of motion $\delta S^{(6)}_f/\delta W_{\mu\nu}=0$ can be easily derived,

\be \p_{\mu}\left(V_{\nu} -\frac 1m \p_{\nu} S\right) + \p_{\nu}\left(V_{\mu} -\frac 1m \p_{\mu} S\right) = m\, \eta_{\mu\nu}\, S \quad . \label{weom}\ee 

\no In the appendix, assuming vanishing fields at infinity, we show that (\ref{weom}) leads to:

\be S=0=V_{\mu} \quad . \label{sv} \ee

 \no Consequently, as in the spin-3 case, the particle content of the deformed model ${\mathcal L}_f^{(6)}$ is the same one of undeformed one ${\mathcal L}_{f=0}^{(6)}$. After the invertible field redefinitions $(h_{\mu\nu\alpha\beta}, A_\mu) \to (\tilde{h}_{\mu\nu\alpha\beta}, \tilde{A}_\mu)$ given by:

\begin{equation}
h_{\mu\nu\alpha\beta} = \tilde{h}_{\mu\nu\alpha\beta} + \frac{\eta_{(\mu\nu}\partial_{\alpha}\tilde{A}_{\beta)}}{3m(D+2)} + \frac{\eta_{(\mu\nu}\partial_{\alpha}\partial_{\beta)}\tilde{h}}{2m^{2}D} \, ,  \label{redefh}
\end{equation}

\begin{equation}
A_{\mu} = \tilde{A}_{\mu} + \frac{3(D+2)}{4mD}\partial_{\mu}\tilde{h} \, ,
\label{redefa}
\end{equation}

\no we can bring (\ref{6thb}) to a second order form, after renaming back
$(\tilde{h}_{\mu\nu\alpha\beta}, \tilde{A}_{\mu}) \to (h_{\mu\nu\alpha\beta}, A_{\mu})$ we have a spin-4 Singh-Hagen model:

\begin{equation}
\begin{aligned}
\mathcal{L}^{(4)}_{f=0} = {}& -\frac{1}{2}(\partial_{\mu}h_{\nu\alpha\beta\rho})^{2} + 2(\partial_{\mu}h^{\mu\nu\alpha\beta})^{2}
+ 6 h_{\mu\nu\alpha\beta}\partial^{\mu}\partial^{\nu}h^{\alpha\beta}  + 3(\partial_{\mu}h_{\nu\alpha})^{2}  + 3(\partial_{\mu}h^{\mu\nu})^{2} \\
& + \frac{3(D^{2}-6D+3)}{4D^{2}}(\partial_{\mu}h)^{2} - \frac{m^{2}}{2}h_{\mu\nu\alpha\beta}^{2} + 3m^{2}h_{\mu\nu}^{2} + \frac{3}{2}m^{2}h^{2} \\
& - \frac{2mD}{(D+2)}h_{\mu\nu}\,\partial^{\mu}A^{\nu} + \frac{m(D^{2}+3D-6)}{D(D+2)}h\,\partial_{\mu}A^{\mu} \\
& + \frac{2D(D+1)}{3(D+2)^{2}}(\partial_{\mu}A_{\nu})^{2} + \frac{D(D-2)}{3(D+2)^{2}}(\partial_{\mu}A^{\mu})^{2} + m^{2}A_{\mu}^{2} \, . \label{l20}  
\end{aligned}
\end{equation}

\no  The equations of motion of (\ref{l20}) lead to $A_{\mu}=0$, the Fierz-Pauli conditions $h_{\mu\nu}=0=(\p\cdot h)_{\mu\nu\alpha}$ and the Klein-Gordon equation $(\Box - m^2)h_{\mu\nu\alpha\beta}=0$ which assures that the f-deformed reduced model (\ref{ms4}) describes 
massive spin-4 particles. The model (\ref{l20}) agrees  with the massive spin-4 theory given in \cite{ady} once we identify their double traceless field $\phi_4$ with $h_{\mu\nu\alpha\beta}-\eta_{(\mu\nu}\eta_{\alpha\beta)}h/(D(D+2))$ and replace the presumably misprinted $-3m$ factor in (4.5) of \cite{ady} by $-2m$.

Next we analyse the massless limit of the spin-3 and spin-4 reduced PBF models.

\section{ The massless limit}

\subsection{ The spin-3 case}

If we take $m\to 0$ in (\ref{lf3m}) we have:

\begin{align}
\mathcal{L}^{(\lambda=3)}_{m\to0}
={}&
\mathcal{L}_{F}(h_{\mu\nu\lambda})
+3\left[
h_{\mu\nu\alpha}\partial^\nu A^\alpha
-h_\mu\Box A^\mu
-\frac{1}{2}h_\mu\partial^\mu(\partial\!\cdot\! A)
\right]
\nonumber\\
&\quad
+\frac{9}{4}(\partial\!\cdot\! A)^2
-f(\partial\!\cdot\! h+\partial\!\cdot\! A)^2
+3\,\mathcal{L}_{EH}(W_{\mu\nu})
+(\partial_\mu z)^2
\nonumber\\
&\quad
+3\,\left[\partial^\mu W-\partial_\nu W^{\mu\nu}\right]\partial_\mu z ,
\label{first} \end{align}

\no where $\mathcal{L}_{EH}$ is the usual  Einstein-Hilbert theory linearized about flat space:

\be \mathcal{L}_{EH} =    \frac{1}{2} W^{\alpha\beta}\, \Box \,  W_{\alpha\beta}+(\del^\mu W_{\mu\beta})^2+  W\, \del^{\mu}\del^{\nu} W_{\mu\nu}-\frac{1}{2} W\,  \Box W \, . \label{leh} \ee

\no After the field redefinitions 

\be W_{\mu\nu} \to  W_{\mu\nu} - \eta_{\mu\nu} \, z /(D-2) \quad ; \quad h_{\mu\nu\alpha} \to h_{\mu\nu\alpha}- \eta_{(\mu\nu}A_{\alpha)}/D \quad . \label{redefwh} \ee

\no followed by $(W_{\mu\nu},A_{\mu}) \to$  $( W_{\mu\nu}/\sqrt{3},-D\,A_{\mu}) $, we achieve a partial diagonalization:

\be
\mathcal{L}^{(3)}_{m\to0}
=
\mathcal{L}^{(3)}_{F}(h_{\mu\nu\lambda}) +\mathcal{L}_{EH}(W_{\mu\nu})
-\frac{3\,D(\, D+1)}{4}F_{\mu\nu}^{\,2}(A)
-f\,(\partial\!\cdot\! h+2\,\partial\!\cdot\! A)^2 
+ \frac{(D+1)}{2(D-2)}\, z\, \Box\, z 
\label{second}
\ee

 At $f=0$ we  have the expected $\pm j$ helicities,  with $j=3,2,1,0$ in $D=4$, but if $f\ne 0$, there is no local field redefinition which diagonalizes the odd spin sector. However, the odd spins 
equations of motion furnish

\begin{equation}
\begin{aligned}
E_{\mu\nu\alpha} \equiv \,\Box h_{\mu\nu\alpha} &- \partial_{(\mu} \partial_\nu h_{\alpha)} - \partial_{(\mu} \partial \cdot h_{\nu\alpha)} 
+ \eta_{(\mu\nu} (\partial \cdot \partial \cdot h)_{\alpha)} - \eta_{(\mu\nu} \Box h_{\alpha)} \\
&- \frac{1}{2} \eta_{(\mu\nu} \partial_{\alpha)} \partial \cdot h + \frac{2f}{3} \eta_{(\mu\nu}\partial_{\alpha)} ( \partial\cdot h + 2 \,\partial \cdot A) = 0 \quad . \label{42}
\end{aligned}
\end{equation}

\begin{equation}
E_\nu \equiv \, \frac{3D(D+1)}{4} \, \partial^\mu F_{\mu\nu}(A) + 2f \, \partial_\nu (\partial \cdot h + 2\, \partial \cdot A) = 0 \label{43}
\end{equation}

\no From $\eta^{\mu\nu} E_{\mu\nu\alpha} = 0$ we have:

\begin{equation}
(\partial \cdot \partial \cdot h)_{\alpha} - \Box\, h_{\alpha} - \frac{\partial_\alpha (\partial \cdot h)}{2} = -\frac{2(D+2)}{3D} f \, \partial_{\alpha} (\partial \cdot h + 2 \,\partial \cdot A) \label{43b}
\end{equation}

\no Substituting back in (\ref{42}) we have

\begin{equation}
\Box h_{\mu\nu\alpha} - \partial_{(\mu} \partial \cdot h_{\nu\alpha)} + \partial_{(\mu} \partial_\nu h_{\alpha)} = \frac{4\,f}{3D} \eta_{(\mu\nu} \partial_{\alpha)}(\partial \cdot h + 2 \,\partial \cdot A) \label{44}
\end{equation}

\no From $\p^{\mu}E_{\mu}=0$ we have $\Box\, (\partial \cdot h + 2 \,\partial \cdot A)=0$, plugging it back in the divergence of (\ref{44}) we derive 

\be \p_{\alpha}U_{\nu} +\p_{\nu} U_{\alpha} = 0  \label{45} \ee

\no which, assuming vanishing fields at infinity, leads to

\be U_{\alpha} = \Box \, h_{\alpha} - (\p\cdot\p\cdot h)_{\alpha} + \frac{\p_{\alpha} (\p\cdot h)}2 - \frac{4\, f}{3\, D}\p_\alpha (\partial \cdot h + 2 \,\partial \cdot A) =0 \label{ualpha} \ee

\no From (\ref{43b}) in (\ref{ualpha}) we obtain for $f\ne 0$, the vanishing of the non diagonal term in (\ref{second})

\be \partial \cdot h + 2 \,\partial \cdot A  =0 \quad . \label{47} \ee

\no Back in (\ref{43}) and (\ref{44}) we have the Maxwell and the spin-3 Fronsdal equations respectively:

\be \p^{\mu}F_{\mu\nu}=0 \quad ; \quad \Box h_{\mu\nu\alpha} - \partial_{(\mu} \partial \cdot h_{\nu\alpha)} - \partial_{(\mu} \partial_\nu h_{\alpha)} = 0 \quad , \label{fronsdal} \ee

All previous equations are invariant under the local transformations:

\be \delta \, h_{\mu\nu\alpha} = \p_{(\mu} \overline{\Lambda}_{\nu\alpha)} \quad ; \quad \delta\, A_{\mu} = \p_{\mu} \, \lambda \quad , \label{50} \ee

constrained by 

\be \p\cdot\p\cdot \overline{\Lambda} + \Box \, \lambda = 0 \quad . \label{51} \ee

\no where $\eta^{\mu\nu} \overline{\Lambda}_{\mu\nu}=0$. Considering (\ref{51}) as a constraint on $\overline{\Lambda}_{\mu\nu}=0$, we can use $\lambda$ in order to fix the usual Lorentz gauge $\p^{\mu}A_{\mu}=0$. Then, (\ref{47}) and (\ref{51}) become respectively $\p \cdot h =0$ and $ \p\cdot\p\cdot \overline{\Lambda} =0$. As explained in \cite{pbf}, those extra conditions define the PBF model which describes $\pm 3$ helicities in $D=4$.  Therefore, the off-shell non diagonal spin-3/spin-1 free theory in (\ref{second}) describes $\pm 3$ and $\pm 1$ helicities. The reader might wonder about the role of the relative factor 2 in (\ref{47}). Similar factors will show up again in the next spin-4 case. We can have a clue on their roles and on possible non local diagonalizing field redefinitions by adding sources. Let us define:

\be {\mathcal{L}^{(3)}_{TJ}} =   \mathcal{L}^{(3)}_{F}(h_{\mu\nu\lambda}) 
-\frac{3\,D(\, D+1)}{4}F_{\mu\nu}^{\,2}(A)
-f\,(\partial\!\cdot\! h+2\,\partial\!\cdot\! A)^2   + h_{\mu\nu\alpha} T^{\mu\nu\alpha}_{PBF} + A_{\mu} J^{\mu} \, . \label{52} \ee  

\no Because of the symmetry (\ref{50}), the sources satisfy weaker constraints as compared to the Fronsdal and Maxwell theories, i.e., 

\bea (\p\cdot T_{PBF})_{\nu\alpha} &=& \eta_{\nu\alpha}\, \Omega + \p_\nu\p_\alpha \, J \quad , \label{53} \\ 
\p^{\mu} J_{\mu} &=& 3 \Box\, J \quad . \label{54} \eea

\no where $\Omega$ and $J$ are arbitrary smooth functions. The usual Fronsdal and Maxwell constraints are recovered if we set $J=0$. 

Supposing that we are allowed to implement non local field redefinitions, we can diagonalize the mixing term in (\ref{52}) in two different ways. First, we can redefine the vector field via a non local $U(1)$ transformation, 

\be A_{\mu} = \tilde{A}_{\mu} - \frac{\p_{\mu}(\p\cdot h)}{2\, \Box} \quad , \label{55} \ee

such that 

\be  h_{\mu\nu\alpha} T^{\mu\nu\alpha}_{PBF} + A_{\mu} J^{\mu}- f\,(\partial\!\cdot\! h+2\,\partial\!\cdot\! A)^2 \to h_{\mu\nu\alpha} T^{\mu\nu\alpha}_{F} + \tilde{A}_{\mu} J^{\mu} - f (2\, \p \cdot \tilde{A})^2 \, , \label{56} \ee 

\no and the new source 

\be (T_F)_{\mu\nu\alpha}   =  (T_{PBF})_{\mu\nu\alpha} - \frac 12 \eta_{(\mu\nu} \p_{\alpha)} J  \quad \label{news} \ee

\no satisfies, due to (\ref{53}), a Fronsdal like non conservation law, i.e., 

\be (\p \cdot T_F )_{\nu\alpha} = \left( \Omega - \frac{\Box J}2 \right) \, \eta_{\nu\alpha}  \quad . \label{58} \ee

\no The generating function now factorizes

\bea {\mathcal{Z}} [T,J] &=& \int {\cal D}h_{\mu\nu\alpha}{\cal D}A_{\mu} e^{\int i\, d^Dx {\mathcal{L}_{TJ}}} \\
&=& \int {\cal D}h_{\mu\nu\alpha} e^{i \int d^D x \left\lbrack {\mathcal{L}_F} (h) + h_{\mu\nu\alpha} T^{\mu\nu\alpha}_{F}\right\rbrack }
{\cal D}\tilde{A}_{\mu} e^{i\int d^D x \,\mathcal{L}_{Max}(\tilde{A},J)} \label{zfact} \eea

\no where
\be \mathcal{L}_{Max}(\tilde{A},J) = - \frac{3D(D+1)}4 F_{\mu\nu}^2(\tilde{A}) - f\, \left(2\, \p\cdot \tilde{A}\right)^2 + \tilde{A}^{\mu} J_\mu \quad . \label{lmaxbj} \ee

\no Using known results for the Fronsdal model two point amplitude we can write 

\be {\mathcal{Z}} [T,J] = e^{i\, \int\, d^Dx\, d^Dy\, \omega(x,y)} \quad , \label{zomega} \ee

\no where 

\be \omega(x,y) = 2 \, \left\lbrack (T_F)_{\mu\nu\alpha} \frac 1{\Box} (T_F)^{\mu\nu\alpha} -\frac 3D (T_F)_{\mu} \frac 1{\Box}(T_F)^{\mu} \right\rbrack + \frac{J^{\mu}[\theta_{\mu\nu}- \omega_{\mu\nu}/\lambda]J^{\nu} }{3\, D(D+1)\Box} \, , \label{wxy} \ee

\no with

\be \omega_{\mu\nu} = \p_\mu\p_{\nu}/\Box \quad ; \quad \theta_{\mu\nu}= \eta_{\mu\nu}-\omega_{\mu\nu}  \quad ; \quad \lambda = \frac{8\, f}{3\, D (D+1)} \quad . \label{wtl} \ee 

Notice that (\ref{wxy}) depends upon the ``gauge parameter'' $\lambda$ because $J^{\mu}$ is not conserved. From (\ref{54}) we can define a transverse current, 

\be J_{\mu}^T \equiv J_{\mu} - 3\, \p_{\mu}J \quad . \label{jmut} \ee

\no Consequently, (\ref{wxy}) becomes, recall $f=3(1-c)/4$,

\be \omega(x,y) = 2 \, \left\lbrack (T_F)_{\mu\nu\alpha} \frac 1{\Box} (T_F)^{\mu\nu\alpha} -\frac 3D (T_F)_{\mu} \frac 1{\Box}(T_F)^{\mu} \right\rbrack + \frac{3\, J^2}{2(c-1)} + \frac{(J_T)^{\mu} (J_T)_{\mu}}{3\, D(D+1)\Box} \quad . \label{wxy2} \ee

\no The second non local way of diagonalizing  (\ref{52}) is by means of a traceless non local Diff,

\be h_{\mu\nu\alpha} = \tilde{h}_{\mu\nu\alpha} + \p_{(\alpha}\overline{\Lambda}_{\mu\nu)}(A) \quad , \label{htil} \ee

\no where

\be\overline{\Lambda}_{\mu\nu)}(A) = \frac{D}{2(1-D)\Box} \left( \p_{\mu} A_\nu + \p_\nu A_{\mu} - \frac 2D \eta_{\mu\nu} \p\cdot A\right) \quad .\label{lambdabar} \ee

\no So we have $\p\cdot h \to \p\cdot \tilde{h} -2\, \p\cdot A $. Since (\ref{htil}) is a symmetry of the Fronsdal theory,  the only change in 
${\mathcal{L}}_{TJ}$ would be:

\be h_{\mu\nu\alpha}T^{\mu\nu\alpha}_{PBF} + A^{\mu} J_{\mu} - f\, (\p\cdot h + 2\, \p\cdot A)^2 \Longrightarrow \tilde{h}_{\mu\nu\alpha}T^{\mu\nu\alpha}_{PBF}  - f\, (\p\cdot \tilde{h})^2 + A_{\mu}J_T^{\mu} \quad . \label{69} \ee

\no Consecutively, the generating function factorizes into the product of the PBF and Maxwell theories:

\be \mathcal{Z}[T,J] = \int {\cal D}\tilde{h}_{\mu\nu\alpha} e^{i\, \int d^Dx\left\lbrack {\mathcal{L}}_{PBF}(\tilde{h}) + h_{\mu\nu\alpha}T^{\mu\nu\alpha}_{PBF}\right\rbrack} \int {\cal D}\tilde{A}_{\mu} e^{i\, \int d^Dx\left\lbrack -\frac 14 F_{\mu\nu}^2(\tilde{A}) + \frac{\tilde{A}_{\mu}\tilde{J}_T^{\mu}}{\sqrt{3D(D+1)}} \right\rbrack}  \label{70}\ee

\no Now comparing (\ref{70}) with (\ref{wxy2}), we read off both, the Maxwell amplitude, last term (\ref{wxy2}), and the spin-3 PBF amplitude. of The explicit calculation of the two point amplitude of the spin-3 PBF model carried out in \cite{epjc26} leads exactly to the terms in (\ref{wxy2}) involving the sources $(T_F^{\mu\nu\alpha}, J)$ with $T_F^{\mu\nu\alpha}$ given in (\ref{news}). Therefore, we recover the full amplitude and (\ref{70}) reproduces (\ref{wxy2}).

Although the diagonalizing non local transformations used here are quite formal steps, our calculations indicate that the consistency of the non diagonal spin-3/spin-1 system given in (\ref{52}) goes beyond the on shell level. The precise numerical factors in (\ref{52}) are just what we need to have propagating $\pm 3$ and $\pm 1$ helicities at quantum level according to the PBF spin-3 model and the Maxwell theory. Notice also that the final output differs from the usual Fronsdal plus Maxwell models by an analytic contact term $(3\, J^2)/(2(c-1))$. Next we analyze the spin-4 case. 

\subsection{ The spin-4 case}

This is the first case where the double traceless condition is required in the corresponding massless theory. The massless limit of the reduced model (\ref{ms4}) splits into even and odd spin theories. 

\be {\mathcal{L}}^{s=4}_{m\to 0} =  {\mathcal{L}}_{{\rm even}} (h_{\mu\nu\alpha},W_{\mu\nu}) +  {\mathcal{L}}_{{\rm odd}} (\chi_{\mu\nu\alpha},B_{\mu}) \quad . \label{71} \ee

\no The odd spin sector coincides, up to an overall normalization factor, with the spin-3/spin-1 system connected via the non diagonal $f$-term as in (\ref{second}) where $h_{\mu\nu\alpha}$ is replaced by $\chi_{\mu\nu\alpha}$, while

\bea  {\mathcal{L}}_{{\rm even}} &=&  {\mathcal{L}}_{F}(h_{\mu\nu\alpha\beta}) + \frac 52 (\p_\mu \, h)^2 - f [(\p\cdot h)_{\mu} + (\p\cdot W)_{\mu} ]^2  \nn\\ &+& 6\left\lbrack W^{\mu\nu}(\p\cdot\p\cdot h)_{\mu\nu} - W^{\mu\nu}\, \Box \, h_{\mu\nu} + (\p\cdot h)^{\mu}(\p\cdot W)_{\mu} - h\, \p\cdot\p\cdot W - \frac 23 h\, \Box \, W \right\rbrack \nn \\
&+& 9(\p\cdot W)_{\mu}^2 - 6 \, W\, \p\cdot\p\cdot W + (\p_\mu W)^2 \quad , \label{72} \eea 

\no where $ {\mathcal{L}}_{F}(h_{\mu\nu\alpha\beta})$ is the spin-4 Fronsdal Lagrangian with however, non vanishing double trace $h=\eta^{\mu\nu}\eta^{\alpha\beta}h_{\mu\nu\alpha\beta}\ne 0$. After a local field redefinition
\be h_{\mu\nu\alpha\beta} \to h_{\mu\nu\alpha\beta}  - \frac 1{(D+2)}\eta_{(\alpha\beta}W_{\mu\nu)}  \quad , \label{hredef4} \ee

\no followed by splitting $h_{\mu\nu\alpha\beta}$ into a double traceless field $\overline{\overline{h}}$ and its double trace $h$ and the redefinitions $W_{\mu\nu} \to W_{\mu\nu} + \eta_{\mu\nu}\,h/(D-2)$ we achieve a partial diagonalization of (\ref{72})

\bea  {\mathcal{L}}_{{\rm even}} &=&  {\mathcal{L}}_{F}(\hdt_{\mu\nu\alpha\beta}) + \frac{6(D+3)}{(D+2)} {\mathcal{L}}_{EH}(W_{\mu\nu}) + \frac{(D+1)(D+3)}{2\, D(D-2)} \varphi \, \Box \varphi \nn\\
&-& f \left\lbrack \p^{\mu}\hdt_{\mu\nu} - \frac{2\, \p_{\nu}\varphi}{D(D-2)} - \frac{2\, \p^{\mu}W_{\mu\nu}}{D+2} - \frac{\p_{\nu}W}{D+2} \right\rbrack^2 \, , \label{73} \eea 

\no where ${\mathcal{L}}_{EH}(W_{\mu\nu}) $ is given in (\ref{leh}) and we have renamed the double trace $h\to \varphi$. 

The equations of motion of (\ref{73}) are given by:

\bea \overline{\overline{E}}_{\mu\nu\alpha\beta} &=& \Box \, \hdt_{\mu\nu\alpha\beta} + \p_{(\mu}\p_\nu \hdt_{\alpha\beta)} + \eta_{(\mu\nu}\p\cdot\p\cdot \hdt_{\alpha\beta} - \eta_{(\mu\nu}\Box\, \hdt_{\alpha\beta)} - \frac 12 \eta_{(\mu\nu}\p_\alpha\p\cdot \hdt_{\beta)} \nn \\ &+& \frac f6 \left\lbrack \eta_{(\mu\nu}\p_\alpha I_{\beta )} - \frac 4D \eta_{(\mu\nu}\eta_{\alpha\beta)} \,\p\cdot I \right\rbrack =0 \, , \label{74} \eea 

\bea E^W_{\mu\nu} &=& \Box \, W_{\mu\nu} - \p_{(\mu}\p\cdot W_{\nu} + \p_\mu\p_\nu W - \eta_{\mu\nu}(\Box\, W - \p\cdot\p\cdot W) \nn\\
&-& \frac{f}{3(D+3)}[\p_{(\mu}I_{\nu)} + \eta_{\mu\nu} \,\p\cdot I ] = 0 \, , \label{75} \eea 

\be (D+1)(D+3) \, \Box \varphi = 4\, f \, \p\cdot I \quad, \label{ephi} \ee

\no where 

\be I_{\rho} = \p^{\mu}\hdt_{\mu\rho} - \frac 2{D(D-2)} \p_{\rho}\varphi -\frac 1{D+2} (\p_{\rho}W + 2\, \p^{\mu}W_{\mu\rho}) \quad . \label{77}\ee

\no From the equations of motion  $\eta^{\alpha\beta}\overline{\overline{E}}_{\mu\nu\alpha\beta}=0$ we have the traceless equation:

\be (\p\cdot\p\cdot \hdt )_{\mu\nu} - \frac 12 \p_{(\mu}(\p\cdot \hdt)_{\nu)}- \Box \, \hdt_{\mu\nu} = - \frac{f\, (D+4)}{6(D+2)}[\p_{(\mu}I_{\nu)} - \frac 2D \eta_{\mu\nu} \,\p\cdot I ] \, . \label{78} \ee

\no Then (\ref{74}) becomes

\be \Box \, \hdt_{\mu\nu\alpha\beta} + \p_{(\mu}\p_\nu \hdt_{\alpha\beta)} - \p_{(\mu}\p\cdot \hdt_{\nu\alpha\beta)} = \frac f{3(D+2)} \left\lbrack \eta_{(\mu\nu}\p_\alpha I_{\beta )} - \frac 4D \eta_{(\mu\nu}\eta_{\alpha\beta)} \,\p\cdot I \right\rbrack , \label{79}\ee

\no From $\p^{\mu}E_{\mu\nu}^W=0$ we have $\Box I_{\mu} = -2 \p_{\mu}(\p\cdot I) $, back in the divergence  of (\ref{79}) we obtain:

\bea \p_{(\nu}K_{\alpha\beta)} &=& 0 \quad , \label{80} \\
K_{\mu\nu} &\equiv & \Box\, \hdt_{\mu\nu} - (\p\cdot\p\cdot\hdt)_{\mu\nu} + \frac12 \p_{(\mu}(\p\cdot\hdt)_{\nu)} - \frac f{3(D+2)}\p_{(\mu}I_{\nu)} \quad . \label{81} \eea 

Supposing vanishing fields at infinity, from (\ref{80}) we can prove, see the appendix, that $K_{\mu\nu}=0$ which altogether with  (\ref{78}) and $f\ne 0$ implies

\be \p_{\mu}\, I_{\nu} + \p_{\nu}\, I_{\mu} = 0 \quad , \label{82} \ee

\no which leads to $I_{\mu}=0$. In summary, our equations of motion is tantamount to:

\bea {\mathcal{F}}_{\mu\nu\alpha\beta}\equiv \Box \, \hdt_{\mu\nu\alpha\beta} + \p_{(\mu}\p_\nu \hdt_{\alpha\beta)} - \p_{(\mu}\p\cdot \hdt_{\nu\alpha\beta)} &=&0   \, , \label{83}\\
\Box W_{\mu\nu} + \p_{\mu}\p_{\nu} W - \p_{(\mu}(\p\cdot W)_{\nu)} - \eta_{\mu\nu} (\Box W - \p\cdot\p\cdot W) &=& 0  \, , \label{84}\\
\Box \, \varphi = 0 \quad ; \quad I_{\mu} &=& 0 \quad , \label{86} \eea

Thus, as in the spin-3 case, the non diagonal term $f\, I_\mu I^\mu$ in (\ref{73}) vanishes on shell. We end up with massless $\pm 4, \pm 2$ helicities and spin-0 equations of motion with the restriction $I_\mu=0$. All equations of motion are invariant under 

\be \delta \hdt_{\mu\nu\alpha\beta} = \p_{(\mu} \overline{\Lambda}_{\nu\alpha\beta)} \quad ; \quad \delta W_{\mu\nu} = \p_\mu \lambda_\nu + \p_{\nu} \lambda_{\mu} \quad , \label{87} \ee

\no such that 

\be (\p\cdot\p\cdot\overline{\Lambda})_{\nu} = \frac{[\Box \, \lambda_{\nu} + 2\, \p_{\nu} (\p\cdot \lambda) ]}{(D+2)} \quad . \label{88} \ee

\no Considering (\ref{88}) as a constraint on $\overline{\Lambda}_{\mu\nu}$, we can use the free parameter $\lambda_{\mu}$ in order to fix the gauge 

\be \p^{\mu}W_{\mu\nu} + \frac{\p_{\nu}W}2 = 0 \quad . \label{89} \ee

\no From the trace of (\ref{84}) we thus have $\Box \, W =0$, so the EH equations of motion (\ref{84}) become 

\be \Box \, W_{\mu\nu} + 2 \p_\mu\p_\nu W = 0 \quad . \label{90} \ee

\no All equations so far are invariant under residual gauge transformations 
$\delta W_{\mu\nu} = \p_{(\mu}\lambda_{\nu)}$ such that $\Box \lambda_{\mu} + 2 \p_{\mu}(\p\cdot \lambda) =0$ which implies $\Box (\p\cdot \lambda)=0$. Since $\delta W = 2 \, \p\cdot \lambda $, we can use the harmonic quantity 
$\p\cdot \lambda $ in order to regauge $W=0$ which finally leads, using (\ref{90}), to $\Box\, W_{\mu\nu} = 0 $. In summary, in the spin-2 sector we have

\be \Box \, W_{\mu\nu} =0 \quad ; \quad \p^{\mu}W_{\mu\nu} = 0 = W \quad . \label{91} \ee 

\no We can further eliminate more $D-1 $ degrees of freedom since all equations are still invariant under transverse harmonic diffs ($\p^\mu \lambda_{\mu}^T=0$ ):

\be \delta W_{\mu\nu} = \p_{(\mu}\lambda_{\nu)}^T \quad  ; \quad \Box \, \lambda_{\mu}^T=0 \quad . \label{92} \ee 

\no At the end $W_{\mu\nu}$ contains the correct number of degrees of freedom for a massless spin-2 particle, i.e.,

\be N_W = \frac{D(D+1)}2 - D - 1 - (D-1) = \frac{D(D-3)}2 \quad . \ee

\no Regarding the spin-4 sector, due to (\ref{91}) the vector condition $I_{\mu}=0$ now becomes

\be \p^{\mu}\hdt_{\mu\nu} = \frac 2{D(D-2)} \p_\nu \varphi \quad . \label{94}\ee

\no From $\eta^{\mu\nu}\p^{\alpha}{\mathcal{F}}_{\mu\nu\alpha\beta}=0 $
we have 

\be 3\, \Box (\p^{\mu}\hdt_{\mu\beta}) - 2\, (\p\cdot\p\cdot\p\cdot \hdt)_{\beta} + \p_{\beta}\, \p^\mu \p^\rho \hdt_{\mu\rho} =0  \quad . \label{95} \ee

\no Due to (\ref{94}) and $\Box \, \varphi =0$ we have $\p^\mu \p^\rho \hdt_{\mu\rho} =0$ and $\, \Box (\p^{\mu}\hdt_{\mu\nu})=0$. Therefore, (\ref{95}) becomes a higher derivative constraint, see also \cite{pbf},

\be (\p\cdot\p\cdot\p\cdot \hdt)_{\mu} = 0 \quad . \label{95b} \ee

Since all equations so far are invariant under $\delta \hdt_{\mu\nu\alpha\beta} = \p_{(\mu}\lbdd_{\nu\alpha\beta)}$ we may fix the de Donder gauge, see \cite{pbf}, 

\be f_{\nu\alpha\beta} = (\p\cdot \hdt)_{\nu\alpha\beta} - \frac 12 \p_{(\nu}\hdt_{\alpha\beta)} = 0 \quad . \label{96} \ee

\no Notice that $\eta^{\nu\alpha} f_{\nu\alpha\beta}=0$ holds identically while  $(\p\cdot \p\cdot f)_{\nu}=0$ follows from previous equations, including (\ref{95b}). So $f_{\nu\alpha\beta}$ has the same number of independent components as $\lbdd_{\nu\alpha\beta}$. From (\ref{96}) and (\ref{83}) we have 

\be \Box \, \hdt_{\mu\nu\alpha\beta} = 0 \quad . \label{97} \ee

\no All previous equation are invariant under the residual harmonic transformations:  $\delta \hdt_{\mu\nu\alpha\beta} = \p_{(\mu}\lbdd_{\nu\alpha\beta)}$ with $\Box\,\lbdd_{\nu\alpha\beta}=0$. Considering (\ref{94}), (\ref{95b}) and (\ref{96}) as well as the residual gauge transformations we end up with the expected number of degrees of freedom for a massless spin-4 particle, i.e., respectively

\bea \!\!\!\!\!\!\!\!N_h &=& \frac{D(D+1)(D+2)(D+3)}{4!} - D - D -2 \!\left\lbrack \frac{D(D+1)(D+2)}{3!} - 2\, D \right\rbrack - \!1 \label{98}\\
&=& \frac{(D-3)(D-2)(D-1)(D+4)}{4\, !} \quad . \label{99} \eea

\no The last factor $-1$ in (\ref{98}) is due to the double traceless condition. Notice that $N_h=2$ in $D=4$ which corresponds to the $\pm 4$ helicities while $N_h=9$ in $D=5$ matches $2\, s +1$ as expected due to the $(D+1)$ (massless)/ $D$ (massive) correspondence.

 Finally,  we have in the massless limit, in $D=4$, all expected  helicities $\pm 4, \pm 3, \pm 2, \pm 1$ and $0$ represented by the fields $\hdt_{\mu\nu\alpha\beta}, \chi_{\mu\nu\alpha} , W_{\mu\nu}, A_{\mu}$ and $\varphi$ respectively.

\section{Conclusion}

We have carried out the Kaluza-Klein (KK) dimensional reduction from $D+1$ to $D$ dimensions of the partially broken Fronsdal (PBF) model suggested in \cite{pbf}. We have worked out the spin-3 and spin-4 cases. The symmetry breaking term with respect to the Fronsdal theory is modulated by the parameter $f$ defined in (\ref{f}). The reduced models describe massive spin-3 and massive spin-4 particles without any extra propagating mode. They differ from  the usual Singh-Hagen (SH) models by perfect square terms as we see in the $f$-dependent terms in (\ref{lint3}) and (\ref{6thb}). We have shown that those $f$-terms vanish on shell, assuring  equivalence of the equations of motion with  the $f=0$ (SH) models.

The triviality of the perfect square terms has its origin in the symmetry breaking term of the massless PBF model (\ref{lc}). It has been argued in \cite{pbf}, by using  Lorentz non covariant gauge invariants, that the content of $\pazocal{L}_{PBF}^{(s)}$ can be read off from the content of the Fronsdal theory $\pazocal{L}_{F}^{(s)}(h)$ and from  $- f\, (\partial \cdot h')^2$ separately as if they were not written in terms of the same field. Since  $- f\, (\partial \cdot h')^2$ has no particle content by itself, the content of $\pazocal{L}_{PBF}^{(s)}$ and  $\pazocal{L}_{F}^{(s)}$ must coincide. The KK reduction on the other hand, is known to preserve the number of degrees of freedom. Indeed, the reduction of $-f \, (\p^MH_M)^2$ gives rise to the square term in (\ref{lint3}) while the reduction of  $-f  (\p^MH_{MN} )^2= -f \, \left\lbrack(\p^MH_{M\mu} )^2+ (\p^MH_{MD} )^2 \right\rbrack$
leads to $-f \, (V_{\mu}^2 + S^2)$ in (\ref{6thb}). By themselves they have no particle content due to the equations of motion of the Stueckelberg fields $A_{\mu} (s=3)$ and $W_{\mu\nu}$ ($s=4$) respectively.

Although the perfect square terms depend upon the physical rank-s field $h$,  it turns out that the  rank-(s-2) Stueckelberg fields $A_{\mu} $ and $W_{\mu\nu}$ only appear in the perfect square terms which assures that the empty content of those $f$-terms is preserved independently of  the rest  of the Lagrangian (SH model). Thus, the argument of \cite{pbf} on the  independence of $\pazocal{L}_{F}^{(s)}(h)$ and  $- f\, (\partial \cdot h')^2$  is somehow confirmed by the dimensional reduction.
Preliminary calculations indicate that the same pattern is true for arbitrary integer spin-s (in progress).

The description in terms of gauge invariant field combinations  as well as the use of ``unitary gauges'', that can be fixed at action level according to \cite{moto}, have naturally lead us to higher derivative massive models, see  (\ref{l34th}) and (\ref{lhz4}) for $s=3$ and (\ref{6th}) and (\ref{6thb}) for spin-4. The spin-3 fourth order model (\ref{lhz4}), after a simple field redefinition $z\to m\, z$, is  the dimensional reduction of  the massless fourth order model of \cite{st}, whose equations of motion had been previously found by J. Schwinger \cite{schw}. Likewise, after field redefinitions, the spin-4 sixth order model (\ref{6thb}) becomes a fourth order model corresponding to the dimensional reduction of the massless fourth order model in formula (45) of \cite{more}.

We have also investigated the peculiar massless limit of the spin-3 and spin-4 cases. In the first case there is no local way of diagonalizing the spin-3 and spin-1 helicities, see (\ref{second}). In the spin-4 case the helicities break up into even and odd spin sectors. The odd spin sector is the same (non-diagonal) one of the spin-3 theory while in the even sector, the helicities $0,2,4$ are intertwined also in a non trivial way, see (\ref{73}). However, the particle content of those spin-3 and spin-4 non diagonal theories still reproduce the massless helicities $\pm j$ with $j=0,1,2,3$ and $j=0,1,2,3,4$ respectively. 

In the massless limit of the spin-3 case, after adding sources, we have successfully reproduced the results of \cite{epjc26} for the  two point amplitude of the PBF model after performing diagonalizing non local field redefinitions. It is amazing how the specific non diagonal terms precisely contribute to reproduce the results of \cite{epjc26}. The amplitude differs from the Fronsdal case by a contact term  which indicates the absence of a map between PBF and Fronsdal that might hold at action level where we have the product of fields at the same spacetime point. So it is unlikely that the equivalence  of the free theories, regarding the propagating modes, will hold true after adding vertices.  We believe that the same pattern goes through for $s>3$.

Finally, we mention that the massive higher spin modes coming from the reduction from $3+1$ to $2+1$ dimensions might play a role in the fractional quantum Hall effect, see \cite{berg}. It would be interesting to investigate the role of the $f$-terms in this regard. 

.
\section{Acknowledgements}

The work of D.D. is partially supported by CNPq  (grant 302635/2026-2), ALR  dos S. is supported by CNPq (grant 102217/2024-6).

\section{Appendix}

Defining $P_{\mu}\equiv V_{\mu} -\frac 1m \p_{\mu} S$, from the trace of (\ref{weom}) we have $S=2\,\partial\cdot{P}/(m\, D)$. Back in (\ref{weom}) leads to
\bea
E_{\mu\nu}= \partial_{\mu}P_{\nu}+\partial_{\nu}P_{\mu}-\frac{2}{D}\eta_{\mu\nu} \partial\cdot{P}\label{A1}
\eea
From $E_{00} = 0$ we obtain
\bea
(D-1)\dot{P}_{0}+\vec\nabla\cdot\vec{P} = 0  \label{A2}
\eea
From $\partial_{i}\partial_{j}E_{ij}=0$ we derive $\nabla^{2}[(D-1)\vec\nabla\cdot\vec{P}+\dot{P}_{0}]=0$
and therefore
\bea
\dot{P}_{0}+(D-1)\vec\nabla\cdot\vec{P}=0 \label{A3}
\eea
From (\ref{A2}) and (\ref{A3}) we have $\dot{P}_{0} = 0 = \vec\nabla\cdot\vec{P}$ and consequently $\partial\cdot{P}=0$
such that $S=0$. Now, from $\partial_{i}E_{ij}=0$ we obtain $P_{j}=0$. From $\partial_{i}E_{i0}=0$ we and up with $P_{0}=0$.
Therefore we demonstrate
\bea
V_{\mu}=0=S \label{A4}
\eea

Now let us prove that from (\ref{80}) we have $K_{\mu\nu}=0$. First we define:
\bea
k_{\mu\nu\alpha} = \partial_{\mu}K_{\nu\alpha}+\partial_{\nu}K_{\alpha\nu}+\partial_{\alpha}K_{\mu\nu}=0 \label{A5}
\eea
Since $\partial^{i}\partial^{j}\partial^{k}k_{ijk}=0$ we have $\nabla^{2}(\partial_{i}\partial_{j}K_{ij})=0$. Then
\bea
\partial_{i}\partial_{j}K_{ij} = 0 \label{A6}
\eea

\no Applying $\partial^{i}\partial^{j}$ in $k_{ijk}$ we have:
$2\nabla^{2}(\partial_{j}K_{jk})+\partial_{k}(\partial_{i}\partial_{j}K_{ij})=0$.
Using (\ref{A6}) we obtain $\nabla^{2}(\partial_{j}K_{jk})=0$ so that
\bea
\partial_{j}K_{jk}=0 \label{A7}
\eea
On the other hand, from  $\partial^{i}k_{ijk} = \nabla^{2}K_{jk}+\partial_{j}\partial_{i}K_{ik}+\partial_{k}\partial_{i}K_{ij}=0$.
and (\ref{A7}) we obtain $\nabla^{2}K_{ij}=0$ which leads to
\bea
K_{jk} = 0 \label{A9}
\eea

\no Applying $\partial^{i}\partial^{j}$ on $k_{0ij}$ we have $\partial^{i}\partial^{j}\dot{K}_{ij}+2\nabla^{2}(\partial_{j}K_{0j})=0$.
Using (\ref{A9}) we obtain $\nabla^{2}(\partial_{j}K_{0j})=0$, thus
\bea
\partial_{j}K_{0j}=0 \label{A10}
\eea

From $\partial^{i}k_{0ij} = \partial^{i}\dot{K}_{ij}+\nabla^{2}K_{0j}+\partial_{j}\partial_{i}K_{0i}=0$, (\ref{A9}) and (\ref{A10}) we find $\nabla^{2}K_{0j}=0$. So
\bea
K_{0j}=0 \label{A11}
\eea
Finally, $k_{00j} = 2\dot{K}_{0j}+\partial_{j}K_{00}=0$ and (\ref{A11}) results $\partial_{i}K_{00}=0 \to K_{00} = 0$. Therefore, we conclude
$K_{\mu\nu}=0$.

\end{document}